\documentclass[twocolumn,showpacs,preprintnumbers,superscriptaddress,prl,amsmath,amssymb]{revtex4}

\usepackage{graphicx}
\usepackage{dcolumn}
\usepackage{bm}

\begin{document}
\title  {Absence of structural transition in (\emph{TM})$_{0.5}$IrTe$_2$ (\emph{TM}=Mn, Fe, Co, Ni)}

\author{J.-Q. Yan}
\affiliation{Materials Science and Technology Division, Oak Ridge National Laboratory, Oak Ridge, Tennessee 37831, USA}
\affiliation{Department of Materials Science and Engineering, University of Tennessee, Knoxville, Tennessee 37996, USA}

\author{B. Saparov}
\affiliation{Materials Science and Technology Division, Oak Ridge National Laboratory, Oak Ridge, Tennessee 37831, USA}

\author{A. S. Sefat}
\affiliation{Materials Science and Technology Division, Oak Ridge National Laboratory, Oak Ridge, Tennessee 37831, USA}

\author{H. Yang}
\affiliation{Department of Materials Science and Engineering, University of Tennessee, Knoxville, Tennessee 37996, USA}
\affiliation{Institute of Materials and Metallurgy, Northeastern University, Shenyang, 110004, PR China}

\author{H. B. Cao}
\affiliation{Quantum Condensed Matter Division, Oak Ridge National Laboratory, Oak Ridge, Tennessee 37831, USA}

\author{H. D. Zhou}
\affiliation{Department of Physics and Astronomy, University of Tennessee, Knoxville, Tennessee 37996, USA}

\author{B. C. Sales}
\affiliation{Materials Science and Technology Division, Oak Ridge National Laboratory, Oak Ridge, Tennessee 37831, USA}

\author{D. G. Mandrus}
\affiliation{Materials Science and Technology Division, Oak Ridge National Laboratory, Oak Ridge, Tennessee 37831, USA}\affiliation{Department of Materials Science and Engineering, University of Tennessee, Knoxville, Tennessee 37996, USA}

\date{\today}

\begin{abstract}
\emph{TM}-doped IrTe$_2$ (\emph{TM}=Mn, Fe, Co, Ni) compounds were synthesized by solid state reaction. Single crystal x-ray diffraction experiments indicate that part of the doped \emph{TM} ions (\emph{TM}=Fe, Co, and Ni) substitute for Ir, and the rest intercalate into the octahedral interstitial sites located in between IrTe$_2$ layers. Due to the lattice mismatch between MnTe$_2$ and IrTe$_2$, Mn  has limited solubility in IrTe$_2$ lattice. The trigonal structure is stable in the whole temperature range 1.80\,K$\leq$\,T\,$\leq$\,300\,K for all doped compositions. No long range magnetic order or superconductivity was observed in any doped compositions above 1.80\,K. A spin glass behavior below 10\,K was observed in Fe-doped IrTe$_2$ from the temperature dependence of magnetization, electrical resistivity, and specific heat. The low temperature specific heat data suggest the electron density of states is enhanced in Fe- and Co-doped compositions but reduced in Ni-doped IrTe$_2$. With the 3\emph{d} transition metal doping the trigonal \emph{a}-lattice parameters increases but the \emph{c}-lattice parameter decreases. Detailed analysis of the single crystal x-ray diffraction data shows that interlayer Te-Te distance increases despite a reduced \emph{c}-lattice. The importance of the Te-Te, Te-Ir, and Ir-Ir bonding is discussed.

\end{abstract}

\maketitle

\section{Introduction}

The correlation between a charge density wave (CDW) state and superconductivity in transition metal dichalcogenides with 1T (octahedrally coordinated) and 2H (trigonal prismatically coordinated) structures has been an interesting topic for decades.\cite{Review} Recently, the intimate interplay in IrTe$_2$ between the charge/orbital density wave, structure/orbital instability, and superconductivity attracted much attention.\cite{CODW,Depolymer,Nohara,Mizokawa,NLWang,Nohara2} The large atomic numbers for Ir and Te imply strong spin-orbital coupling (SOC). IrTe$_2$ thus offers a material platform to investigate structure/charge/orbital fluctuations and superconductivity under strong SOC. The formation of an orbital Peierls state has been suggested to drive the structural transition, with orbital fluctuations mediating low temperature superconductivity. A study of the orbital physics in IrTe$_2$ may provide clues to understanding the pairing mechanism in iron pnictide superconductors where orbital effects are being debated.

\begin{figure} \centering \includegraphics [width = 0.47\textwidth] {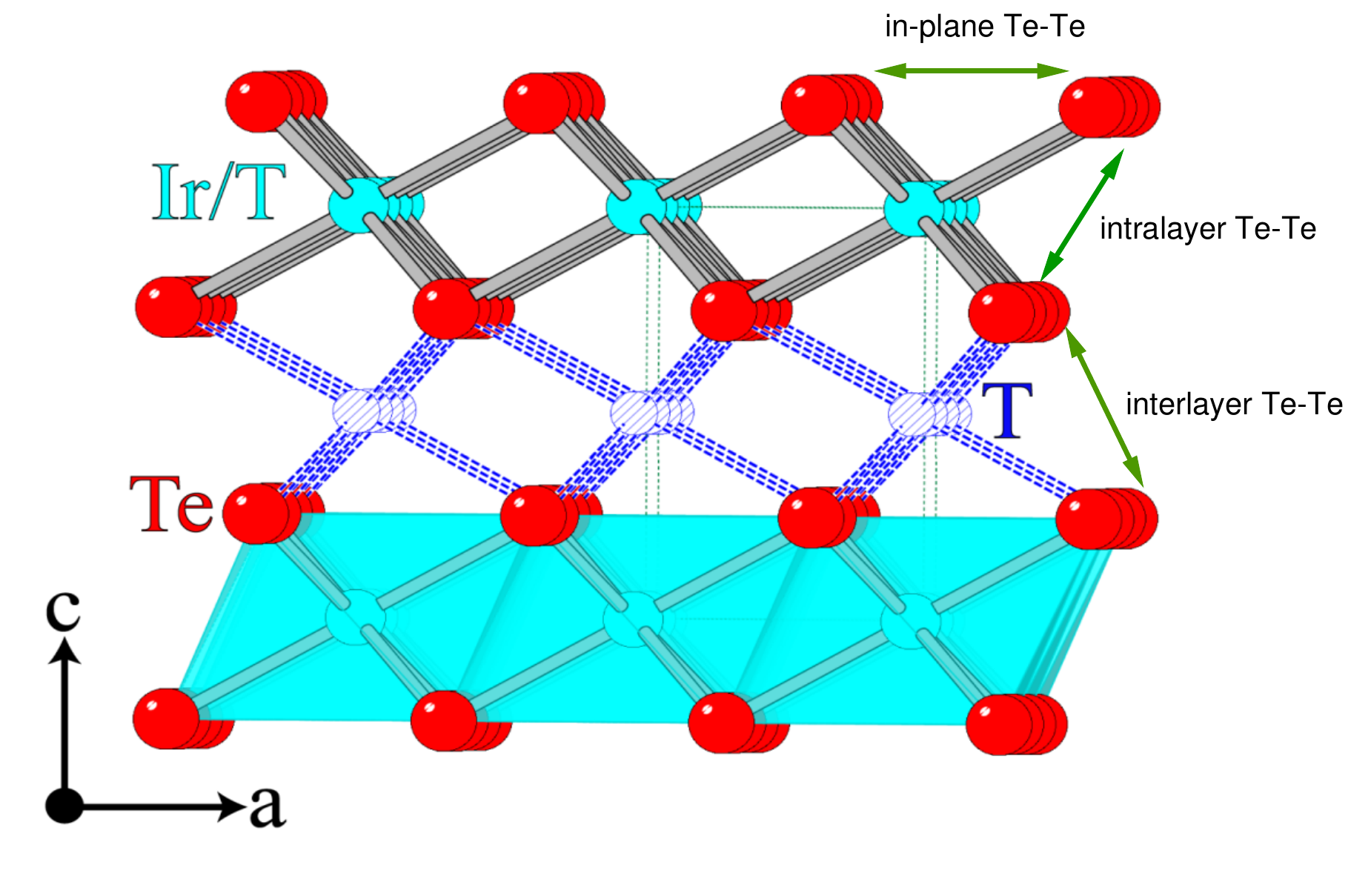}
\caption{(color online) Crystal structure of IrTe$_2$ and two possible sites for doped \emph{TM} ions: the Ir substitutional site, and the intercalational site in between IrTe$_2$ sheets. Different Te-Te interatomic distances are specified. See text for more details.}
\label{Figure-1}
\end{figure}

At room temperature, IrTe$_2$ crystallizes in the polymeric CdI$_2$-type structure (space group \emph{P-}3\emph{m}1, see Fig.\,\ref{Figure-1}) with short Te-Te interatomic distance. A structural phase transition to a low temperature commensurately modulated triclinic structure (space group \emph{P}1) takes place at $\sim$280\,K accompanied by distinct anomalies in transport, magnetic, thermodynamic, and optical properties.\cite{Huibo,NLWang} Several mechanisms have been proposed to drive the structural transition: formation of a charge/orbital density wave or orbital Peierls state, kinetic energy, and Te 5\emph{p} bonding instabilities.\cite{CODW,Mizokawa,NLWang,Huibo} Despite much effort, the origin of the structural transition is still under debate.

The structural transition takes place at a higher temperature under hydrostatic pressure or when Te is partially substituted by Se.\cite{Depolymer,Haidong} In contrast, the transition is usually suppressed once foreign transition metal ions are introduced into the lattice. As illustrated in Fig.\,\ref{Figure-1}, transition  metal dopants can occupy the octahedral sites in-between the IrTe$_2$ layers, i.e., the intercalation site, or the Ir site, i.e., the substitutional site. Previous studies showed that doping transition metal ions at either site suppresses the structural transition.\cite{CODW,Nohara,Nohara2,Rh,CuxIrTe2} With the suppression of the structural transition, superconductivity emerges and appears to compete with the former.

When Cu is intercalated into Cu$_x$IrTe$_2$, the structural transition disappears and superconductivity emerges at \emph{x}$\sim$0.03 with T$_c$$\sim$3.0\,K.\cite{CuxIrTe2} T$_c$ shows little variation with \emph{x} up to 0.10. At \emph{x}\,=\,0.50, an anomaly in the temperature dependence of electrical resistivity and magnetization is observed at T$\approx$250\,K suggesting the reappearance of the structural transition.\cite{CuIr2Te4JPCS, CuIr2Te4JLTP} This doping effect is similar to that in \emph{TM}$_x$TiSe$_2$ compounds.\cite{MxTiSe2} TiSe$_2$ shows a structural phase transition induced by a CDW below $\sim$200\,K. Similar to IrTe$_2$, the structural transition is accompanied by distinct anomalies in transport, magnetic and thermodynamic properties. Intercalation of 3d transition metals in \emph{TM}$_x$TiSe$_2$ (\emph{TM}=Cr, Mn, Fe, Ni) suppresses the structural transition which disappears at \emph{x} $\sim$0.1. However, when \emph{x} $\geq$ 0.25, pronounced resistivity anomalies characteristic of the updoped compound are observed suggesting the reappearance of the CDW state. Local distortions in Se-Ti-Se layer, which are affected by the intercalated dopants, are believed to switch on/off the CDW state. The possible reappearance of the structural transition in Cu$_{0.5}$IrTe$_2$ deserves more study to (1) provide experimental evidence, such as from x-ray and/or neutron diffraction, for the structural transition at T$\approx$250\,K, (2) unravel the underlying mechanism that drives the structural transition and compare with that in IrTe$_2$, and (3) explore the evolution of the structural transition and the electronic ground state in the composition range 0.10$\leq$x$\leq$0.50. It's also of great interest to study whether the structure transition reappears in other \emph{TM}$_{0.5}$IrTe$_2$ (\emph{TM}=Mn, Fe, Co, and Ni) compounds.

Our synthesis effort failed to obtain either single phase Cu$_{0.5}$IrTe$_2$ powder or single crystals with the correct composition. CuTe is observed as the impurity in polycrystalline samples prepared by conventional solid state reaction of a stoichiometric mixture of Cu, Ir and Te powders following the procedure reported in Ref\cite{CuIr2Te4JLTP,CuIr2Te4JPCS}. We also tried to intercalate Cu by long term annealing the mixture of Cu and IrTe$_2$ below 300$^o$C.\cite{CuxIrTe2}  Annealing at 280$^o$C for a month still doesn't reach a complete intercalation which suggests this is limited by the diffusion kinetics. Single crystal growth was performed starting with various charge/flux ratio in different metallic fluxes. No Cu$_{0.5}$IrTe$_2$ single crystals were found to grow out of Te, Cu-Te, or Bi flux. Cu and Te mixture melts at rather low temperatures in a wide composition range which makes crystal growth and powder synthesis difficult.\cite{PD}

In this work, compounds with the nominal composition \emph{TM}$_{0.5}$IrTe$_2$ (\emph{TM}=Mn, Fe, Co, and Ni) were synthesized and investigated by x-ray powder and single crystal diffraction, magnetization, electrical resistivity and specific heat. Our results show that (1) Mn has a limited solubility in IrTe$_2$. Fe, Co, and Ni are observed at both the Ir site and the intercalation site;(2) the trigonal structure is stable in the whole temperature range 1.8\,K\,$\leq$\,T\,$\leq$\,300\,K studied; (3) no superconductivity was observed above 1.80\,K; and (4) compared to the parent compound, the doped compositions have a larger \emph{a} lattice parameter but a reduced value for \emph{c}. The reduced \emph{c} lattice parameter is accompanied by a shortened intralayer Te-Te distance and an increased interlayer Te-Te separation.

\begin{figure} \centering \includegraphics [width = 0.47\textwidth] {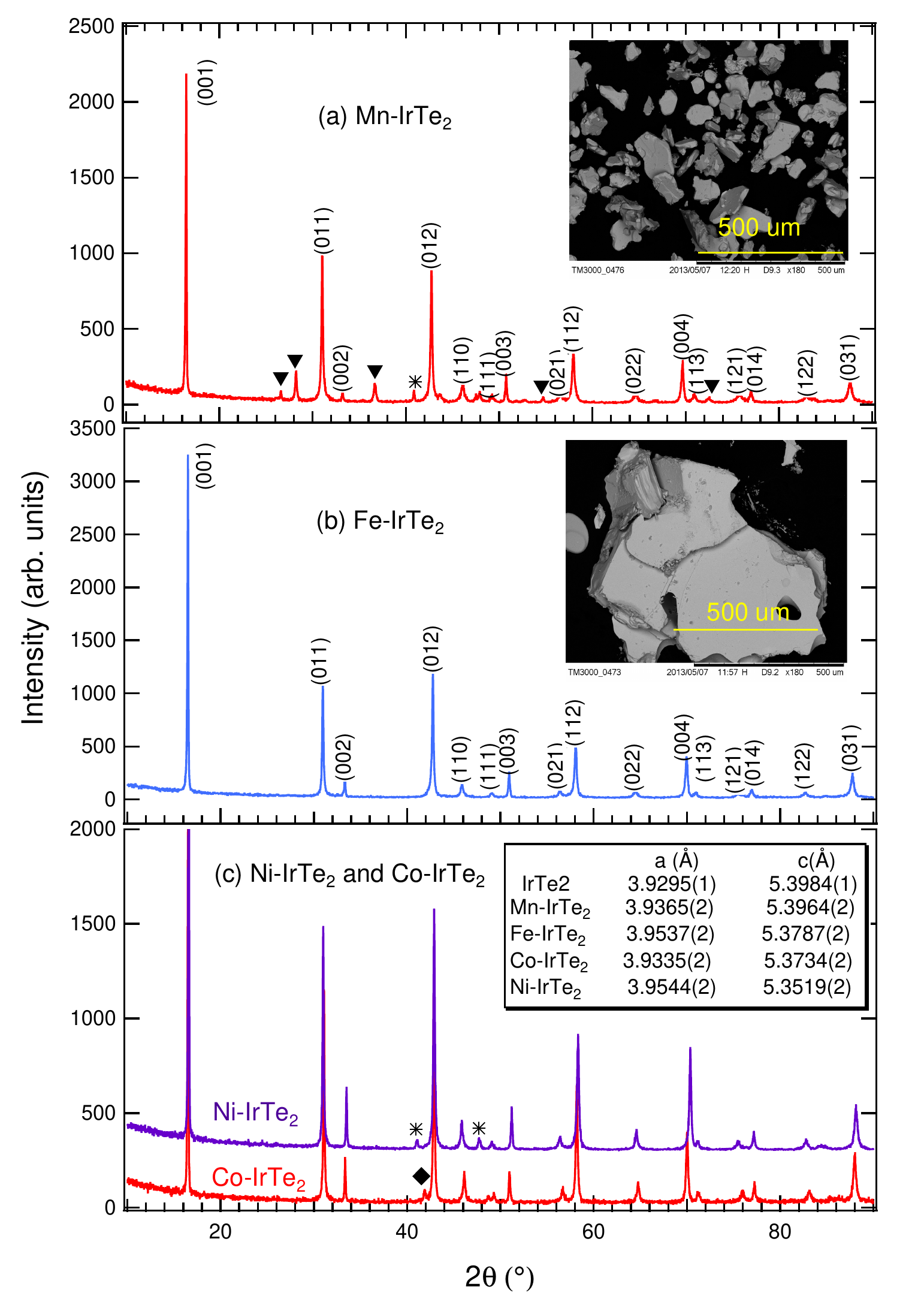}
\caption{(color online) X-ray powder diffraction patterns of \emph{TM}-doped IrTe$_2$: (a) Mn-IrTe$_2$, (b) Fe-IrTe$_2$, and (c) Ni-IrTe$_2$ and Co-IrTe$_2$. Reflections from impurities are indicated by $\blacktriangledown$ (MnTe), $\blacklozenge$ (unknown), and $\ast$(Ir). Insets in (a) and (b) show the SEM images of the micron sized plate-like crystals. The lattice parameters are summarized in inset of (c). Lattice parameters for IrTe$_2$ are from Ref\cite{Huibo}.}
\label{XRD-1}
\end{figure}

\section{Experimental Details}
Polycrystalline materials were synthesized via a conventional solid state reaction method. The starting powders, Ir (Alfa, 99.99\%), \emph{TM}=Mn, Fe, Co, Ni (Alfa, 99.9999\%), and Te (Alfa, 99.999\%), were mixed in the atomic ratio of \emph{TM}:Ir:Te\,=\,1:2:4. All metal powders were reduced at 800$^o$C for 12 hours in the forming gas of Ar+4\%H$_2$ before using. After a thorough mixing inside of a dry glove box, the powder was pelletized and transferred to an alumina crucible. The alumina crucible was then sealed in a quartz tube backfilled with 1/3 atmosphere of high purity Ar. The sealed ampoule was heated to 1000$^o$C over 12 hours, held at 1000$^o$C for 120 hours, and then cooled to room temperature over 15 hours.

Room temperature X-ray diffraction patterns were collected on a X'Pert PRO MPD X-ray Powder Diffractometer using the Ni-filtered Cu-K$_\alpha$ radiation.  The X'Pert HighScore Plus software was employed to identify possible phases and determine the lattice parameters. After sintering, the pellets contain numerous plate-like crystals. To study the morphology and composition of these plates, Scanning Electron Microscope (SEM) imaging and energy-dispersive X-ray spectroscopy were carried out using a Hitachi-TM3000 microscope equipped with a Bruker Quantax 70 EDS system. Data acquisition was carried out with an accelerating voltage of 15 kV and a scanning time of 2 min.

The plate-like crystals are large enough for single crystal x-ray diffraction study. The diffraction measurements were performed on a Bruker SMART APEX CCD-based single crystal X-ray diffractometer with Mo K$_\alpha$ ($\lambda$ = 0.71073\,\emph{${\AA}$}) radiation. Crystals were selected under an optical microscope and cut to a suitable size (0.1\,mm on all sides) inside Paratone N oil. Because the crystals are very soft and malleable, extreme care was taken not to deform them. The crystals were then cooled to 173(2) K using a cold nitrogen stream and X-ray intensity data were collected at this temperature. The structure solution by direct methods and refinement by full matrix least-squares methods on \emph{F}$^2$ were carried out using the SHELXTL software package.\cite{software} SADABS was used to apply absorption correction.

Magnetic properties were measured with a Quantum Design (QD) Magnetic Properties Measurement System in the temperature interval 1.8\,K\,$\leq$\,T\,$\leq$\,300\,K. The temperature dependent specific heat and electrical transport data were collected using a 9 Tesla QD Physical Properties Measurement System in the temperature range of 1.9\,K\,$\leq$\,T\,$\leq$\,300\,K.

\section{Experimental Results}

Significant grain growth takes place after sintering at 1000$^o$C. Plate-like crystals (see insets of Fig.\ref{XRD-1}) could be observed in fired pellets for all compositions. The pressed pellets become rather porous and weak after sintering due to the weak connections between the plate-like crystals. The largest dimension of Fe-doped IrTe$_2$ plates can be up to 0.6\,mm. The plates are smaller for other compositions with the largest dimension approximately 0.3\,mm. X-ray powder diffraction patterns were collected at room temperature for all compositions. For Fe-doped IrTe$_2$, all reflections in the diffraction pattern can be indexed with \emph{P-}3\emph{m}1 symmetry. Weak reflections of free Ir were observed in the diffraction patterns of Ni-IrTe$_2$. There is one weak reflection that cannot be identified in the diffraction pattern of Co-IrTe$_2$. By contrast, as shown in Fig \ref{XRD-1}\,(a), a larger fraction of free Ir and MnTe were observed as impurities in Mn-IrTe$_2$. A Mn-IrTe$_2$ pellet fired at 1000$^o$C for 120\,hours was reground, pelletized, and fired at the same temperature for another 120\,hours. Unfortunately, the refiring doesn't reduce the amount of impurities suggesting that Mn has limited solubility in IrTe$_2$. In all diffraction patterns collected, the strong (00\emph{l}) reflections imply that plate-like crystals are formed after sintering due to grain growth.

The room temperature lattice parameters are summarized in the inset of Fig.\ref{XRD-1}(c). Data for the parent compound are from Ref.\cite{Huibo}. Compared with the parent compound, all doped compositions exhibit larger \emph{a}-axis but smaller \emph{c}-axis lattice parameter. Mn-IrTe$_2$ shows a small deviation of the lattice parameters from IrTe$_2$, which suggests a limited doping of Mn in IrTe$_2$ and agrees with the observation of MnTe and Ir impurities in the powder diffraction pattern. Ni-IrTe$_2$ shows  the largest reduction of the \emph{c}-axis; while for Co-IrTe$_2$, a significant reduction of the \emph{c}-axis is accompanied with a small change of the \emph{a}-axis.

\begin{figure} \centering \includegraphics [width = 0.47\textwidth] {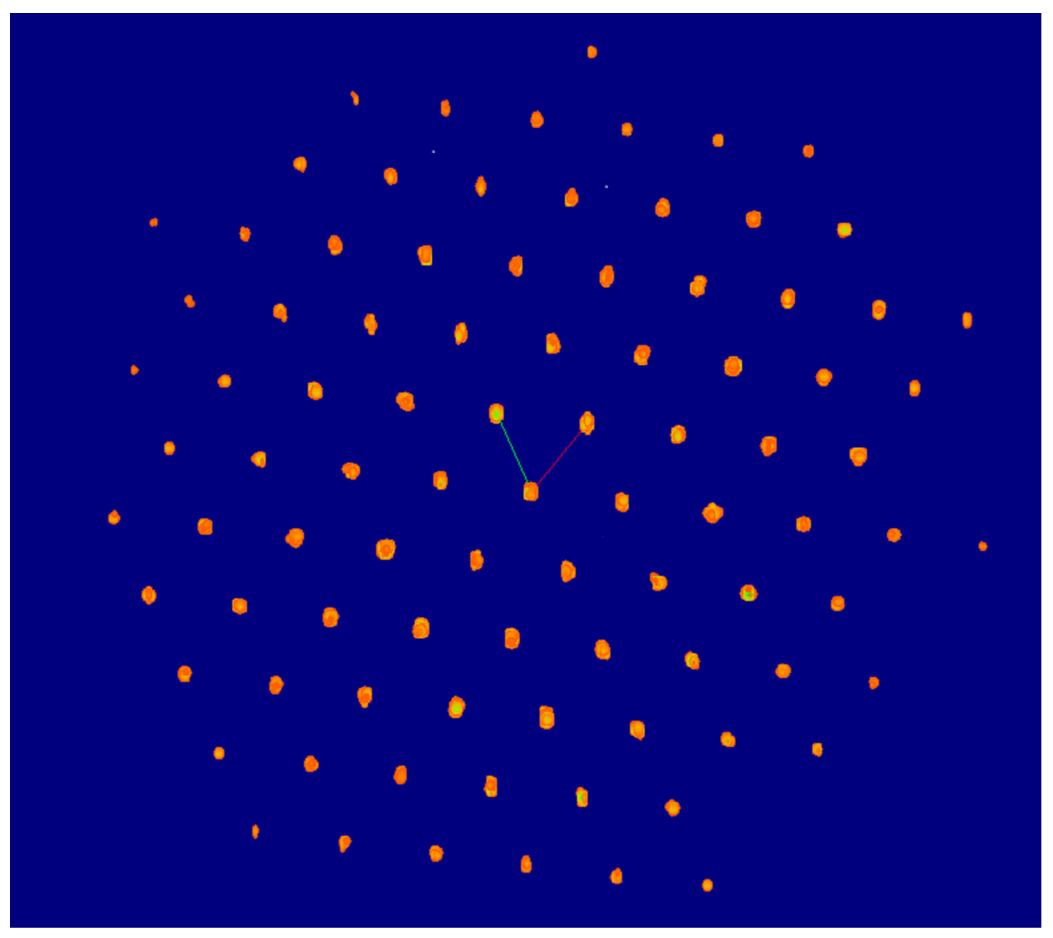}
\caption{(color online) Single crystal x-ray diffraction pattern of Fe-IrTe$_2$ along the reciprocal \emph{c}-axis taken at 173\,K. Red and green lines show the reciprocal \emph{a}- and \emph{b}-axis, respectively.}
\label{SXRD-1}
\end{figure}

The small plate-like crystals enable single crystal x-ray diffraction study. Figure\,\ref{SXRD-1} shows the single crystal x-ray diffraction pattern taken at 173\,K for Fe-IrTe$_2$ as an example. All peaks could be indexed with the \emph{P-}3\emph{m}1 symmetry; no superlattice peaks were observed at 173\,K. Room temperature structural data from Ref\cite{Huibo} for IrTe$_2$ were used to get an initial refinement. For Mn-doped IrTe$_2$ crystal, this refinement gave good R-factors, and a featureless final difference Fourier map, indicating that the Mn content is below the detection limit. For Fe-/Co-/Ni-doped crystals, R-values were elevated with R$_1$ $\approx$ 0.048/0.052/0.064 and \emph{w}\emph{R}$_2$ $\approx$ 0.130/0.125/0.173. Additionally, the difference Fourier maps showed large residual difference peaks of 8.70\,e$^-$/{\AA}$^3$, 12.52\,e$^-$/{\AA}$^3$, and 17.99\,e$^-$/{\AA}$^3$ for Fe-, Co-, and Ni- doped crystals, respectively, located at the interstitial position 2.64\,${\AA}$ away from Te. Further refinements indicated that the interstitial positions are partially occupied by the transition metal atoms, and that these elements also partially substitute for Ir in the layers. In the final model, we included the interstitial site and also refined the occupation at the Ir position. A summary of single crystal x-ray diffraction data and refinement parameters for \emph{TM}-doped IrTe$_2$ crystals is provided in Table 1. Atomic coordinates and equivalent isotropic displacement parameters are listed also in Table 2. The doping dependence of lattice parameters agrees with that determined from room temperature x-ray powder diffraction patterns.

\begin{table*}[!ht]
\caption{Selected crystallographic data and refinement parameters for Fe$_{0.33(2)}$Ir$_{0.83(1)}$Te$_2$, Co$_{0.36(1)}$Ir$_{0.78(1)}$Te$_2$, Ni$_{0.42(2)}$Ir$_{0.78(1)}$Te$_2$, and Mn$_x$Ir$_{1-y}$Te$_2$. \emph{x} and \emph{y} in Mn$_x$Ir$_{1-y}$Te$_2$ are below the detection limit as described in text.\\}
\label{table1}
\begin{ruledtabular}
\begin{tabular} {llllll}
Empirical formula	&Fe$_{0.33(2)}$Ir$_{0.83(1)}$Te$_2$& Co$_{0.36(1)}$Ir$_{0.78(1)}$Te$_2$ & Ni$_{0.42(2)}$Ir$_{0.78(1)}$Te$_2$ & Mn$_x$Ir$_{1-y}$Te$_2$\\
Formula weight&	433.16&	426.33	&429.19 &N/A\\
Temperature (K)	& 173(2)& 173(2)&173(2)&173(2)\\
Radiation, wavelength (\emph{${\AA}$})&Mo K$\alpha$, 0.71073&Mo K$\alpha$, 0.71073&Mo K$\alpha$, 0.71073 &Mo K$\alpha$, 0.71073\\
Space group, \emph{Z}	&\emph{P-}3\emph{m}1 (No. 164), 1&\emph{P-}3\emph{m}1 (No. 164), 1&\emph{P-}3\emph{m}1 (No. 164), 1 &\emph{P-}3\emph{m}1 (No. 164), 1\\
\emph{a} (${\AA}$)	&3.9540(4)	&3.9336(3)&	3.9536(4)&	3.9394(7)\\
\emph{c} (${\AA}$)	&5.3694(11)&	5.3614(10)&	5.3405(12)&	5.3844(17)\\
\emph{V} (${\AA}$$^3$)&	72.70(2)&	71.84(2)	&72.29(2)&72.36(3)\\
Calculated density (g/cm$^3$)&	9.894	&9.854	&9.858&N/A\\
Absorption coefficient (mm$^{-1}$)&	59.072&	57.878&	58.113&65.53\\
$\theta$ range (º)&	3.79-28.12&	3.80-28.27&	3.82-28.14&	3.78-28.27\\
\emph{R}$_1$* (all data)	&0.0306&	0.0186	&0.0239&0.0254\\
\emph{wR}$_2$* (all data)&	0.0837&	0.0496&	0.0717&0.0634\\
Goodness-of-fit on F$^2$	&1.252&	1.178	&1.264&1.304\\
Largest diff. peak/hole (e$^-$/${\AA}$$^3$)&	1.64/-3.16&	1.17/-2.04	&2.57/-1.91&2.601/-2.673\\
\end{tabular}
\end{ruledtabular}
* \emph{R}$_1$=$\sum$$|$$|$F$_0$$|$-$|$F$_c$$|$$|$/$\sum$$|$F$_0$$|$; \emph{w}\emph{R$_2$}=$|$$\sum$$|$\emph{w}(F$_0$$^2$-F$_c$$^2$)$^2$$|$/$\sum$$|$\emph{w}(F$_0$$^2$)$^2$$|$$|$$^{1/2}$, where \emph{w}=1/$|$$\sigma$$^2$F$_0$$^2$+(AP)$^2$+BP$|$, and P=(F$_0$$^2$+2F$_c$$^2$)/3; A and B are weight coefficients.
\end{table*}

\begin{table*}[!ht]
\caption{Atomic coordinates, equivalent isotropic displacement parameters (\emph{U$_{eq}$$^a$}) and site occupation factors (SOF) for Fe$_{0.33(2)}$Ir$_{0.83(1)}$Te$_2$, Co$_{0.36(1)}$Ir$_{0.78(1)}$Te$_2$, Ni$_{0.42(2)}$Ir$_{0.78(1)}$Te$_2$, and Mn$_x$Ir$_{1-y}$Te$_2$. \emph{x} and \emph{y} in Mn$_x$Ir$_{1-y}$Te$_2$ are below the detection limit as described in text.\\}
\label{table2}
\begin{ruledtabular}
\begin{tabular} {llllllll}
Atom&	Wyckoff site &	\emph{x}	&\emph{y}	&\emph{z}	&\emph{Ueq} ($\AA$$^2$)	&SOF\\
\hline
Fe$_{0.33(2)}$Ir$_{0.83(1)}$Te$_2$\\
Ir/Fe1&	1a&	0&	0&	0&	0.0231(6)&	0.83(1)/0.17(1)\\
Te	&2d&	1/3	&2/3	&0.7488(2)	&0.0247(5)	&1\\
Fe2&	1b	&0	&0	&1/2	&0.038(8)&	0.16(2)\\
\\						
Co$_{0.36(1)}$Ir$_{0.78(1)}$Te$_2$\\
Ir/Co1&	1a&	0&	0&	0&	0.0154(3)&	0.78(1)/0.22(1)\\
Te	&2d	&1/3	&2/3	&0.7502(1)&	0.0167(3)&	1\\
Co2&	1b&	0&	0&	1/2	&0.023(5)&	0.137(11)\\
\\						
Ni$_{0.42(2)}$Ir$_{0.78(1)}$Te$_2$\\
Ir/Ni1&	1a	&0	&0&	0&	0.0168(5)&	0.78(1)/0.22(1)\\
Te	&2d	&1/3&	2/3	&0.7495(1)	&0.0176(5)&	1\\
Ni2	&1b	&0	&0	&1/2&	0.031(5)&	0.198(16)\\
\\						
Mn$_x$Ir$_{1-y}$Te$_2$\\
Ir&	1a	&0	&0&	0&	0.0170(5)&	1\\
Te	&2d	&1/3&	2/3	&0.7476(2)	&0.0179(5)&	1\\
\end{tabular}
\end{ruledtabular}
*\emph{U$_{eq}$} is defined as one third of the trace of the orthogonalized \emph{U$_{ij}$} tensor.
\end{table*}

\begin{figure} \centering \includegraphics [width = 0.47\textwidth] {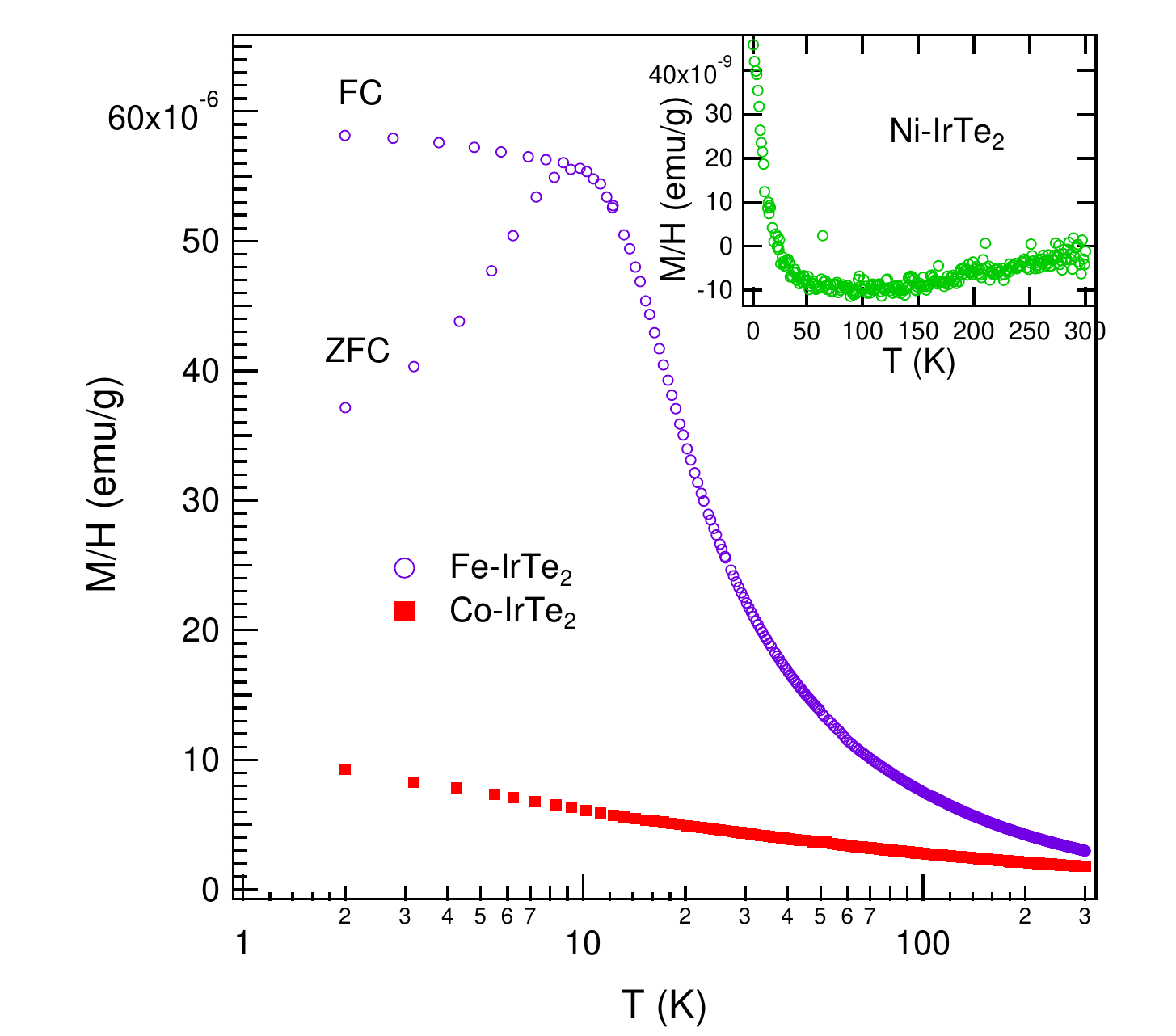}
\caption{(color online) The temperature dependence of magnetization measured in an applied magnetic field of 1\,kOe for the samples as indicated.}
\label{MT-1}
\end{figure}

The temperature dependence of the magnetization was measured in an applied magnetic field of 1\,kOe in both zero-field-cooled (ZFC) and field-cooled (FC) modes and is shown in Fig. \ref{MT-1}. A splitting between ZFC and FC curves was observed for Fe-IrTe$_2$ below 10\,K, which is a sign of long range magnetic order at T$_N$=10\,K or spin glass behavior with T$_f$=10\,K. For \emph{TM}-IrTe$_2$ (\emph{TM}=Co, and Ni), FC and ZFC curves overlap in the whole temperature range with a Curie-Weiss tail at low temperatures likely due to isolated magnetic impurities. For Mn-IrTe$_2$ (not shown), the MnTe impurity phase contributes to the magnetization, and a slope change was observed at $\sim$310\,K where long range magnetic order occurs in MnTe.\cite{MnTe} For all four compositions studied, no rapid drop of magnetization similar to that reported in IrTe$_2$ or Cu$_{0.5}$IrTe$_2$ was observed. The absence of this drop and a lack of hysteresis suggest that the trigonal structure is stable down to 1.8\,K in \emph{TM}-IrTe$_2$ (\emph{TM}\,=\,Mn, Fe, Co, and Ni). This statement is further supported by electrical resistivity and specific heat measurements.

\begin{figure} \centering \includegraphics [width = 0.47\textwidth] {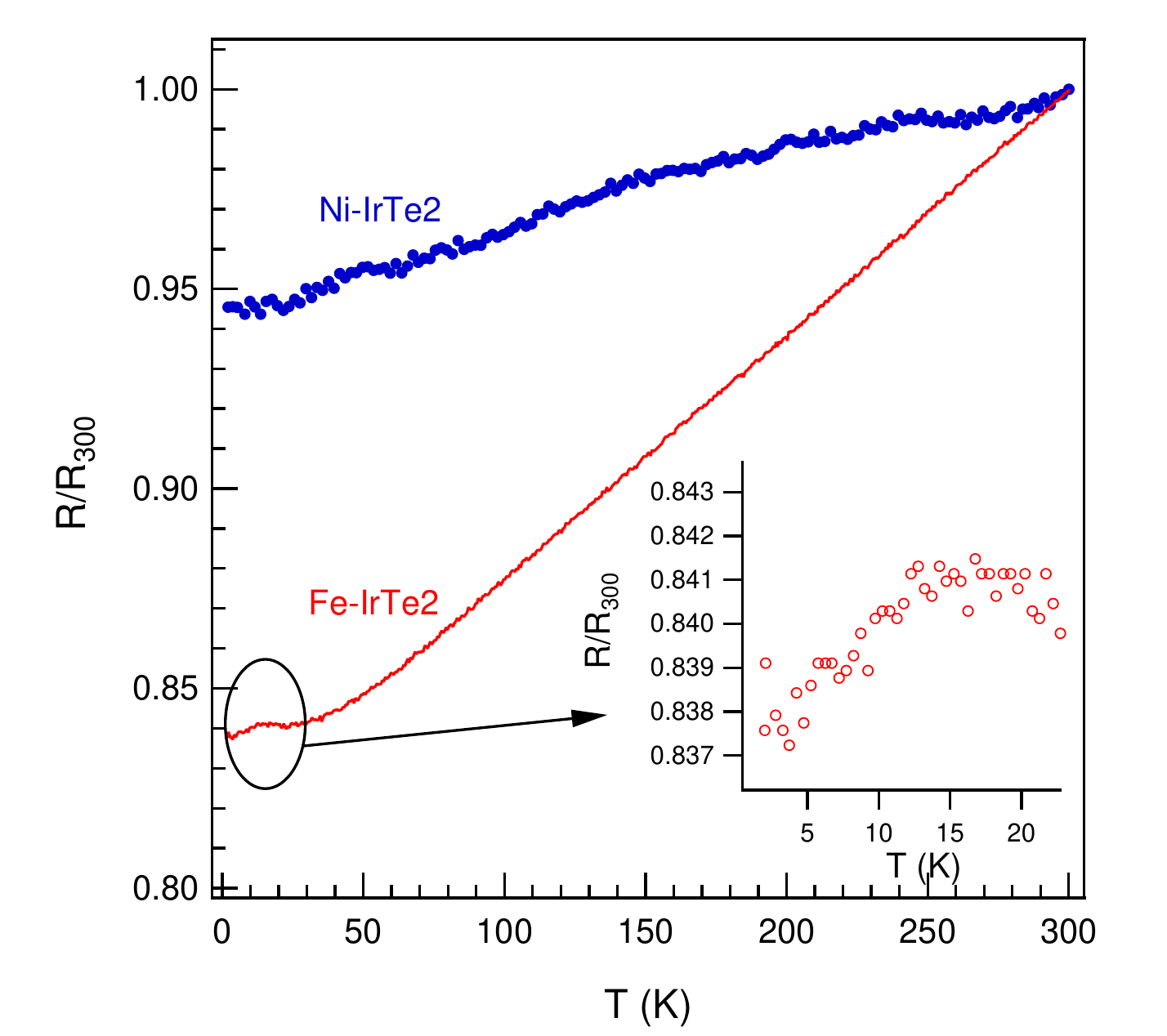}
\caption{(color online) Temperature dependence of electrical resistivity normalized to the value at 300\,K. Inset highlights the drop around 10\,K in Fe-IrTe$_2$.}
\label{RT-1}
\end{figure}

Figure\,\ref{RT-1} shows the normalized electrical resistivity of Fe-IrTe$_2$ and Ni-IrTe$_2$. One crystal of Fe-IrTe$_2$ with the largest dimension of 0.5\,mm was used in the study. A porous rectangular bar was used to measure the resistivity of Ni-IrTe$_2$ and the serious grain boundary scattering leads to scattered data. No transport data were obtained for Co-IrTe$_2$ because the fired pellet was too weak and the single crystal plates were too small.  As shown in Fig.\,\ref{RT-1}, a metallic behavior was observed in the whole temperature range with no evidence of a structural phase transition. No superconductivity was observed above 1.90\,K. The electrical resistivity curves measured on both heating and cooling overlap and no hysteresis was observed. For Fe-IrTe$_2$, a slight drop was observed around 10\,K which suggests reduced scattering associated with the magnetic feature around 10\,K.

\begin{figure} \centering \includegraphics [width = 0.47\textwidth] {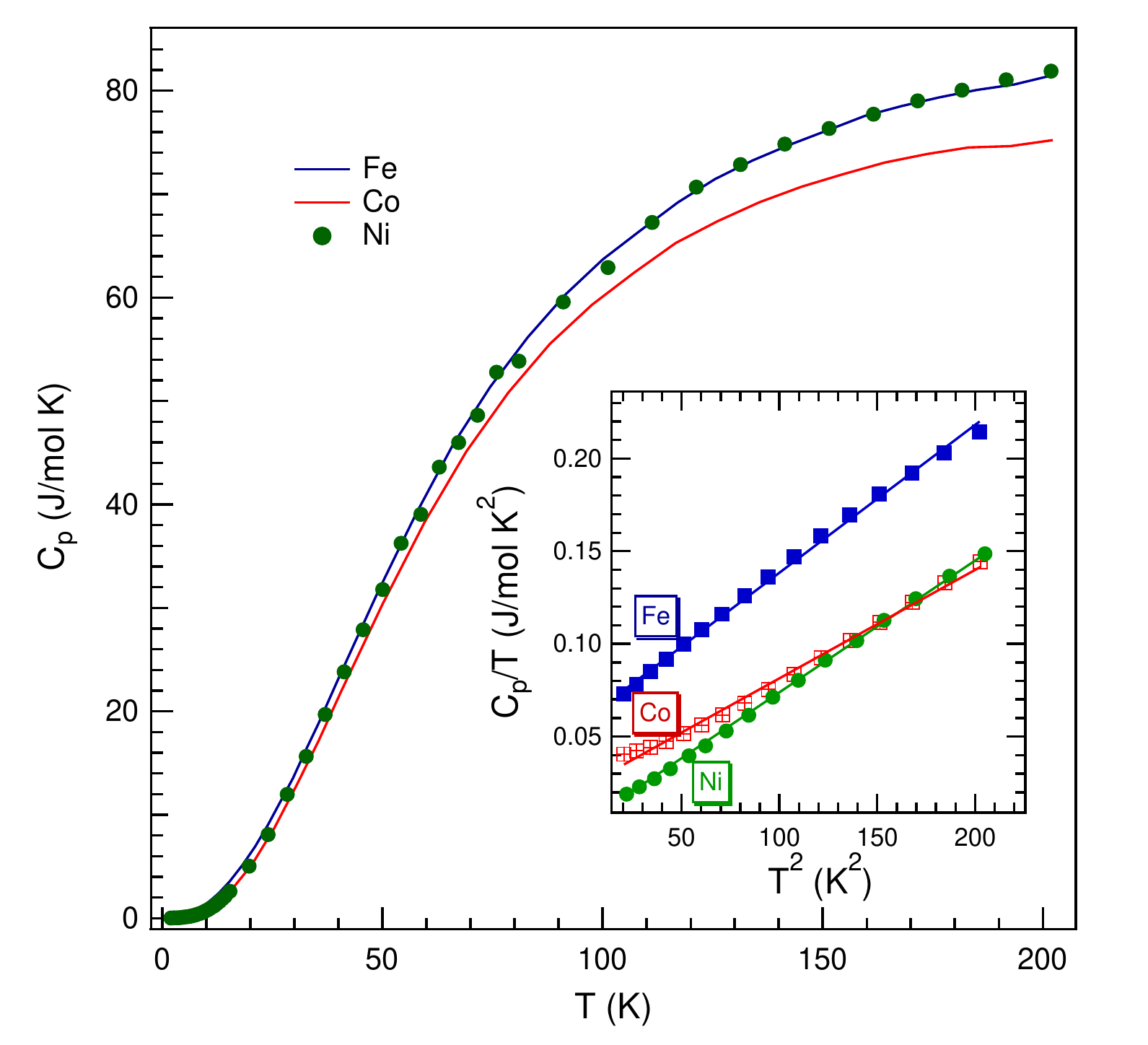}
\caption{(color online) The temperature dependence of specific heat of \emph{TM}-IrTe$_2$ (\emph{TM}=Fe, CO, and Ni). Inset shows the linear fitting of Cp/T vs T$^2$.}
\label{Cp-1}
\end{figure}

Figure\,\ref{Cp-1} shows the temperature dependence of specific heat measured in the temperature range 1.9\,K$\leq$T$\leq$200\,K. No sign of the structural transition was observed in the studied temperature range. No anomaly was observed at 10\,K for Fe-IrTe$_2$ even though an anomaly was observed in the temperature dependence of magnetization and electrical resistivity. These results suggest a spin glass behavior for Fe-IrTe$_2$ below 10\,K, which is further confirmed by $\mu$SR measurements.\cite{Luke}

As shown in the inset, the specific heat follows the relation C$_p$/T=$\gamma$+$\beta$T$^2$ at low temperatures. The linear fitting yields $\gamma$=54(1), 21(1), and 1.2(4) mJK$^{-2}$mol$^{-1}$ for Fe-IrTe$_2$, Co-IrTe$_2$, and Ni-IrTe$_2$, respectively. In Ir$_{1-x}$Pt$_x$Te$_2$, the trigonal structure is stabilized to the lowest temperature when \emph{x}$\geq$0.04 and $\gamma$ decreases from $\sim$6\,mJK$^{-2}$mol$^{-1}$ with increasing \emph{x}. The much larger $\gamma$ coefficient for Fe-IrTe$_2$ and Co-IrTe$_2$ reported in this study suggests much larger electronic density of states (DOS) at the Fermi level. The Debye temperature can be estimated using the equation $\beta$=(12N$_A$$\pi$$^4$n\emph{k}$_B$)/(5$\Theta$$_D$$^3$), where n is the number of atoms per formula unit, N$_A$ is Avogadro constant and \emph{k}$_B$ is Boltzmann constant. $\Theta$$_D$ is 431\,K, 484\,K, and 460\,K for Fe-IrTe$_2$, Co-IrTe$_2$, and Ni-IrTe$_2$, respectively.

\section{Discussion}

As shown in Fig.\ref{Figure-1}, there are two crystallographic positions for dopant transition metal ions. At both positions, the transition metal ions stay in the center of octahedra with similar distortions. Our x-ray diffraction studies show that Fe, Co and Ni take both positions with significant solubilities. The Mn content in Mn-IrTe$_2$ is below the detection limit of our single crystal x-ray diffraction. However, there is a minor Mn inclusion in the structure. First, the structural transition that occurs at $\sim$280\,K for IrTe$_2$ is absent in Mn-IrTe$_2$. Second, a x-ray powder diffraction study observed a large fraction of Ir and MnTe impurities. And finally, the lattice parameters of Mn-IrTe$_2$ determined from both powder and single crystal diffraction deviate little from those of IrTe$_2$. To understand the solubility of different dopants in IrTe$_2$, one would intuitively compare the structure of \emph{TM}Te$_2$ and IrTe$_2$ and the octahedral distortion.

NiTe$_2$ has a hexagonal, polymeric CdI$_2$-type structure as does IrTe$_2$. The lattice parameters and corresponding Ir-Te bond lengths are about 2$\sim$3\% smaller than those of IrTe$_2$. FeTe$_2$ and CoTe$_2$ exhibit an orthorhombic marcasite structure.\cite{FeTe2} The FeTe$_6$ octahedron in FeTe$_2$ has one short Fe-Te bond (2.508(8)\,${\AA}$),  one intermediate (2.5589(7)\,${\AA}$), and one long bond (2.635(8)\,${\AA}$). The average Fe-Te bond length is 2.567\,${\AA}$. The CoTe$_6$ octahedron has two long bonds both about 2.602(11)\,${\AA}$ and one short bond of 2.5815(6)\,${\AA}$, with an average of 2.595\,${\AA}$. These bonds are 0.5$\sim$5\% shorter than those in IrTe$_2$ with an average Ir-Te bond length of 2.650\,${\AA}$. The above structural similarities and/or the smaller \emph{TM}Te$_6$ octahedra might account for the substantial amount of dopants in Fe-, Co-, and Ni-IrTe$_2$. In contrast, MnTe$_2$ crystallizes in a pyrite-type primitive cubic structure, in which Mn-Te bond is 2.908\,${\AA}$. The large difference of the crystal structure and/or the octahedra volume between MnTe$_2$ and IrTe$_2$ might lead to the limited solubility of Mn in IrTe$_2$. Despite the limited solubility, the absence of any anomaly in the temperature dependence of magnetization and specific heat (not shown) implies that the structure transition in IrTe$_2$ is sensitive to Mn doping and Mn stabilizes the high temperature trigonal phase.

\begin{figure} \centering \includegraphics [width = 0.47\textwidth] {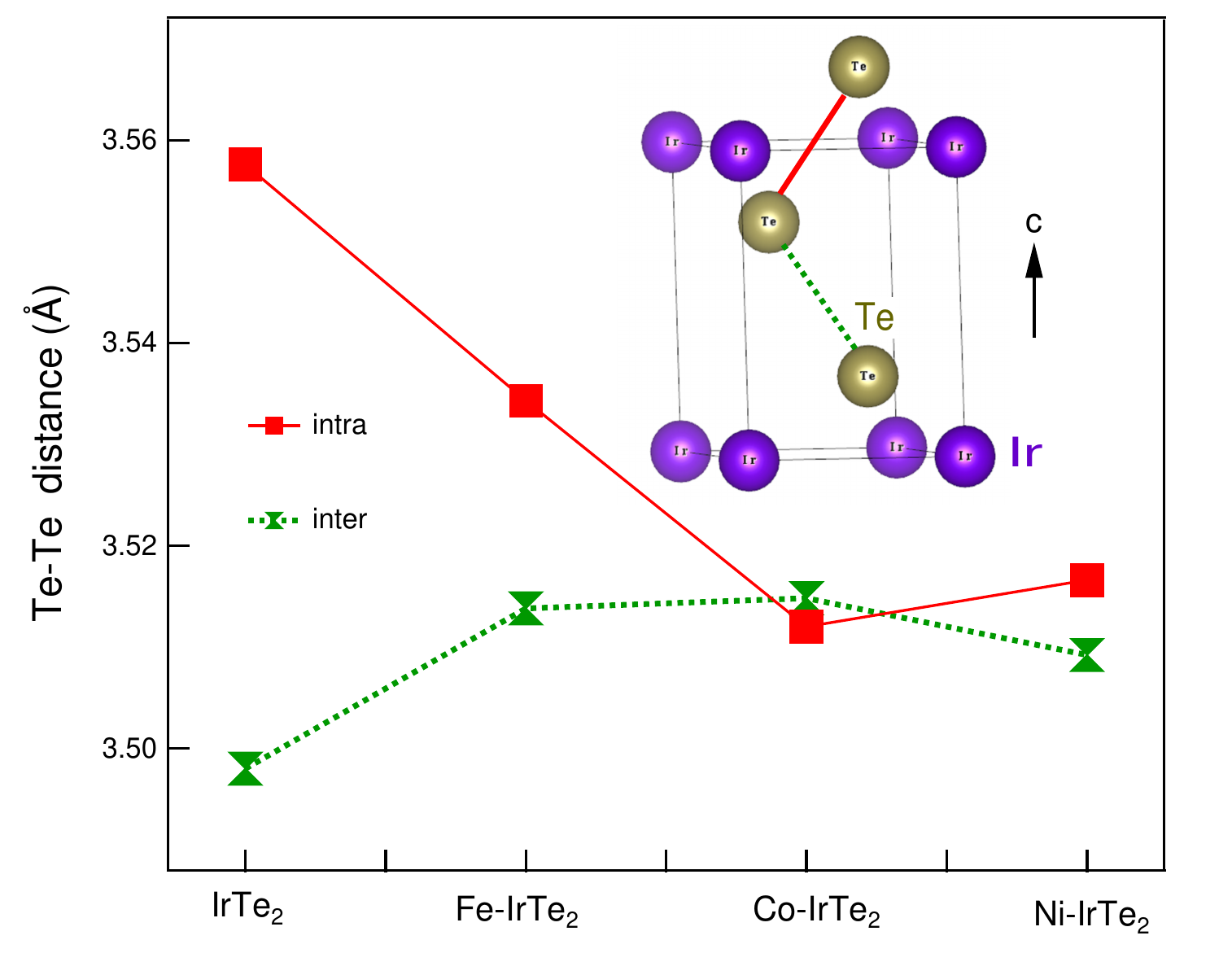}
\caption{(color online) The intralayer and interlayer Te-Te distances in \emph{TM}-doped IrTe$_2$. Inset shows the Te-Te distances described in the figure. The solid and dashed curves are guide for the eyes.}
\label{bonds-1}
\end{figure}

The structure transition in IrTe$_2$ is sensitive to hydrostatic pressure and chemical doping at both Ir and Te sites. Hydrostatic pressure up to 2\,GPa stabilizes the low-temperature phase.\cite{Depolymer,Haidong} The low temperature phase has a smaller volume which provides an intuitive view of the pressure effect. Hydrostatic pressure was argued also to increase the ratio of the interlayer/intralayer Te-Te distances, which depolymerizes the polymeric Te-bond network. The same mechanism is proposed to account for the depolymerization effect of partial substitution of Te by Se. Figure\,\ref{bonds-1} shows the interlayer and intralayer Te-Te distances in IrTe$_2$ (300\,K) and doped compositions (173\,K). For all doped compositions, the ratio of interlayer/intralayer Te-Te distances is larger than that of the parent compound. However, our x-ray and other physical property measurements suggest the trigonal (\emph{P}-3\emph{m}1) phase is stable and no structure transition was observed in the temperature range investigated. Obviously, the depolymerization effect itself cannot explain the stabilization of the trigonal phase in our doped compositions. This might be due to the significant amount of intercalation doping.

Since our dopants take both possible crystallographic positions shown in Fig.\,1, it might be informative to check the effect of dopants at each site on the physical properties and crystal structure. The relation between the high temperature structural transition and low temperature superconductivity has been the focus of various studies: Ir$_{1-x}$Pd$_x$Te$_2$ (0$\leq$x$\leq$0.10),\cite{CODW} Ir$_{1-x}$Pt$_x$Te$_2$ (0$\leq$x$\leq$0.25),\cite{Nohara} Ir$_{1-x}$Rh$_x$Te$_2$ (0$\leq$x$\leq$0.30),\cite{Rh} Pd$_x$IrTe$_2$ (0$\leq$x$\leq$0.10),\cite{CODW}  Cu$_x$IrTe$_2$ (0$\leq$x$\leq$0.10).\cite{CuxIrTe2} The electronic phase diagrams of all previously studied systems are similar: the structural transition is suppressed with doping and disappears; followed by superconductivity with T$_c$ up to 3\,K. With respect to the lattice parameters, the \emph{c}-axis expands with the intercalation of Pd; but no obvious change was observed when Cu is intercalated or Ir is partially substituted by Rh; other dopants suppress the \emph{c}-axis. For the \emph{a}-axis, Rh substitution induces little modification; for other dopants, it always increases with doping regardless of which site the dopants occupy. The lattice parameter change in our doped compositions agrees well with the above observations by showing an increased \emph{a}-axis but a reduced \emph{c}-axis. As described above and illustrated in Fig.\ref{bonds-1}, the \emph{c}-lattice parameter contraction with doping reduces the intralayer Te-Te distance but increases the interlayer one. Since the increased interlayer/intralayer Te-Te distance ratio cannot account for the absence of a structural transition in this work, we examine the change in the \emph{a}-lattice parameter associated with the reduced intralayer Te-Te distance.

Both PdTe$_2$ and PtTe$_2$ have the polymeric CdI$_2$ structure with a larger \emph{a} but a smaller \emph{c} than IrTe$_2$.\cite{PdTe2} The doping dependence of lattice parameters in Ir$_{1-x}$Pd$_x$Te$_2$ and Ir$_{1-x}$Pt$_x$Te$_2$ suggests that Vegard law is observed. NiTe$_2$ has the same CdI$_2$ structure as does IrTe$_2$ but with smaller lattice parameters. One would expect a reduced \emph{a}- and \emph{c}-lattice parameter if Vegard's law is observed. Thus the observed larger \emph{a}-lattice parameter in this study is abnormal.

The lattice parameter \emph{a} corresponds to the in-plane Ir-Ir and Te-Te interatomic distance. The larger \emph{a}- lattice reduces the in-plane Ir-Ir interactions and Te-Te overlap. A recent band structure calculation with the newly determined triclinic structure suggests that the structure transition is due to a local bonding instability associated with the Te 5\emph{p} states.\cite{Huibo} The reduced intralayer Te-Te separation might enhance the local bonding instability due to the enhanced overlap repulsion. The stabilization of the high temperature trigonal phase in our doped compositions suggests that increasing in-plane Ir-Ir and Te-Te spearation might relieve the bonding instabilities. On the other hand, an orbitally induced Peierls mechanism was proposed to be responsible for the structural transition and associated resistivity as well as optical anomalies, and Fermi surface reconstruction. \cite{Nohara,Mizokawa,Mizokawa2} The increased Ir-Ir separation prevents the formation of an orbital Peierls state and maintain the orbital fluctuations which have been suggested to mediate superconductivity. Our results cannot distinguish between the above scenarios. The observed bond length change and the stabilization of the trigonal phase, however, highlight the importance of the orbital hybridization in IrTe$_2$-based materials especially with a doping induced reduction of the intralayer Te-Te distance.

With an electronic ground state configuration of 5\emph{s}$^2$5\emph{p}$^4$, the nominal valence state of Te can vary from -2 to +6 depending on the degree of covalency in different transition metal tellurides. Due to the extended Ir 5d shell and the similar electronegativity, a large orbital hybridization is expected between Ir and Te in IrTe$_2$. The calculation of the orbital overlap population at the Fermi level suggests very covalent bonding between Ir and Te.\cite{Jobic1991} The strong covalency can induce the electron transfer, which would affect the effective charge state of both Ir and Te. The charge balance of IrTe$_2$ has been suggested to be Ir$^{3+}$(Te$^{-1.5}$)$_2$.\cite{PdTe2,Jobic1991} The effective charge for Fe, Co, and Ni with similar distorted octahedral coordination is proposed to be +3, +3, and +4, respectively.\cite{MeTe2} With this simplified scenario for the effective charge state and with a rigid band model, Fe$^{3+}$ doped in IrTe$_2$ will dope holes into the valence band and lower the Fermi level. This leads to a larger electronic DOS, as manifested by the specific heat data. On the other hand, the much narrower 3\emph{d} bands at the Fermi level can also contribute to a large electronic DOS. However, the above ionic picture cannot explain the DOS change in Co- and Ni-doped compositions. This, in turn, suggests the importance of orbital hybridization on the electronic properties of IrTe$_2$-based materials. Band structure calculations are needed to examine the effects of dopants at each crystallographic site.

We noted that partial substitution of Ir by Rh leads to little change of lattice parameters.\cite{Rh} RhTe$_2$ also crystallizes in polymeric CdI$_2$ structure with lattice parameters and an octahedral distortion similar to those in IrTe$_2$. This small structural difference accounts for the weak dependence of lattice parameters on doping. Despite the close similarity of the structures, the electron transfer between Te and Rh/Ir would be different since Ir has more extended 5\emph{d} orbitals. This covalency difference leads to a modification of the electronic structure with increased Rh doping, and thus the suppression of the structural transition and the emergence of superconductivity.

\section{Conclusions}
\emph{TM}$_{0.5}$IrTe$_2$ (\emph{TM}=Mn, Fe, Co, and Ni) compounds were synthesized by sintering the starting elements at 1000$^o$C. Mn has a limited solubility in IrTe$_2$ possibly due to the large structure difference between IrTe$_2$ and MnTe$_2$. Doped transition metal ions occupy both the Ir site and the intercalation site. No superconductivity was observed down to 1.80\,K in any of the doped compositions. X-ray diffraction, magnetization, electrical resistivity, and specific heat measurements show that doping stabilizes the trigonal structure in the entire temperature range(1.9\,K-300\,K), in contrast to the reappearance of the structural transition in Cu$_{0.5}$IrTe$_2$. Compared to the parent compound, doped compositions have a larger \emph{a}-lattice parameter but a reduced \emph{c}. Doping increases the interlayer Te-Te distance. The smaller \emph{c}-lattice parameter comes from the reduction of the intralayer Te-Te separation, which increases the overlap of the Te 5\emph{p} bands with the Ir 5\emph{d} bands. The increasing \emph{a}-lattice parameter with doping reduces the in-plane Ir-Ir and Te-Te overlap, which seems to stabilize the trigonal phase. Our results suggest that both the strong hybridization between Te 5\emph{p} and Ir 5\emph{d} orbitals and the effect of in-plane Ir-Ir and Te-Te interactions should be considered in future experimental and theoretical efforts to understand the origin of the structural transition and low-temperature superconductivity in IrTe$_2$-based materials.

The structural transition reappears in \emph{TM}$_x$TiSe$_2$ (\emph{TM}=Cr, Mn, Fe, and Ni) and Cu$_{0.5}$IrTe$_2$ where doped transition metal ions were intercalated. The observation that \emph{TM} ions occupy both the interstitial and intercalation sites in \emph{TM}$_{0.5}$IrTe$_2$ (\emph{TM}=Mn, Fe, Co, and Ni) and the absence of any structural transition in these compositions imply different effects for substitutional and intercalation dopants that should be studied in future experiments.

\section{Acknowledgments}
JQY thanks Minghu Pan for helpful discussions. BIS and HBC thank R. Custelcean for his assistance with the single crystal X-ray diffraction measurements. Work at ORNL was supported by the U.S. Department of Energy, Basic Energy Sciences, Materials Sciences and Engineering Division.

\end{document}